\documentclass{llncs}
\usepackage{makeidx}
\usepackage{amsmath, amssymb}
\usepackage{listings}
\usepackage{color}
\usepackage{courier}
\usepackage{tikz}
\usepackage{braket}
\usepackage{tabularx}
\usepackage[subtle]{savetrees}
\usetikzlibrary{automata,arrows}
\definecolor{lightgray}{rgb}{.5,.5,.5}
\definecolor{darkgray}{rgb}{.3,.3,.3}

\lstdefinelanguage{JavaScript}{
  keywords={typeof, new, true, false, catch, function, return, null, catch, switch, var, if, in, while, do, else, case, break},
  keywordstyle=\color{black}\bfseries,
  ndkeywords={class, export, boolean, throw, implements, import, this},
  ndkeywordstyle=\color{darkgray}\bfseries,
  identifierstyle=\color{black},
  sensitive=false,
  comment=[l]{//},
  morecomment=[s]{/*}{*/},
  commentstyle=\color{lightgray}\ttfamily,
  stringstyle=\color{darkgray}\ttfamily,
  morestring=[b]',
  morestring=[b]",
  moredelim=[is][stringstyle]{@<}{>@},
}

\title{Model Checking Regular Language Constraints}
\author{Arlen Cox \and Jason Leasure}
\institute{IDA/CCS}
\begin{document}
\maketitle

\begin{abstract}
  Even the fastest SMT solvers have performance problems with regular expressions from real programs. Because these performance issues often arise from the problem representation (e.g. non-deterministic finite automata get determinized and regular expressions get unrolled), we revisit Boolean finite automata, which allow for the direct and natural representation of any Boolean combination of regular languages.  By applying the IC3 model checking algorithm to Boolean finite automata, not only can we efficiently answer emptiness and universality problems, but through an extension, we can decide satisfiability of multiple variable string membership problems. We demonstrate the resulting system's effectiveness on a number of popular benchmarks and regular expressions.
\end{abstract}

\section{Introduction}

While there are a significant number of satisfiability-modulo-theories (SMT) solvers for strings~\cite{z3str2,cvc4,norn,hampi}, most of the effort in these solvers has gone into solving word equations where the primary problem is string concatenation.  This paper is concerned with the problem of SMT solving where theory predicates are regular expression membership.  For example the following formula should be determined to be valid:
\begin{align*}
  \forall x \in \textrm{STRINGS}. \; \left( x \in (\texttt{a} \cdot \texttt{b})^* \vee x \in (\texttt{c} \cdot \texttt{d} \cdot \texttt{e})^* \right) \rightarrow x \in (\texttt{c} \cdot \texttt{d} \cdot \texttt{e} \ | \ \texttt{a} \cdot \texttt{b} )^*
\end{align*}
In this case, it must be proven that any string that consists of a sequence of ``ab'' repeated arbitrarily many times must be in the set of strings generated by repeating any combination of ``ab'' or ``def'' arbitrarily many times.

These kinds of problems arise in analysis of a variety of systems ranging from structured configuration files to general purpose programming languages.  For example, the key challenge in symbolic execution~\cite{king} of the program shown in Listing~\ref{lst:discount} is determining possible values for the string \lstinline|input|.  When \lstinline|student_discount| is true, strings must be email addresses because of the first regular expression, must end in \lstinline|.edu| because of the second regular expression, and when \lstinline|student_discount| is false strings must either not be email addresses or must not be email addresses that end in \lstinline|.edu|.

\begin{figure}
  \centering
\begin{lstlisting}[
    caption=Determine discount with regular expressions,
    label=lst:discount
  ]
if(@</^[a-z]+@[a-z]+\.[a-z]+$/>@.test(input)) {
  if(@</\.edu$/>@.test(input)) {
    student_discount = true;
  } else {
    student_discount = false;
  }
} else {
  student_discount = false;
}
if(student_discount) {
  // discount code
}
\end{lstlisting}
\end{figure}

A traditional way of handling regular expressions is to convert them to non-deterministic finite automata (NFA)~\cite{thompson:1968} and then use automata operations to combine them.  For instance, to construct an automaton whose acceptance of \lstinline|input| coincides with \lstinline|student_discount| being false, one constructs the union of the complement of the NFA representing first regular expression with the complement of the NFA for the second regular expression.  The problem with this is that complementation is performed via the powerset construction, which can result in an exponential increase in the number of states in the automaton.  As a result solving simple problems can quickly become intractable.

Boolean finite automata (BFA)~\cite{bfa} provide a \textit{lazy} representation of these operations.  Like NFA, there is a direct translation from regular expressions, but unlike NFA, Boolean finite automata also support complementation and intersection operations without an increase in the size of the automaton.  This means that any given SMT problem may be exponentially more compact when expressed using BFA rather than NFA.  The tradeoff is in worst-case computational complexity.  The worst case for NFA emptiness testing is linear while BFA emptiness testing is PSPACE-complete.  However, empirically we find that hardware model checking algorithms such as IC3~\cite{ic3} can be applied to BFA and are particularly effective for emptiness checking.

When using a hardware model checker to decide BFA emptiness, the hardware model checker learns lemmas to quickly trim large portions of the search space.  This often results in proofs that avoid the exponential cost of conversion to an NFA.  Similarly, example strings that are accepted by a BFA can often be found efficiently.

In exploring the benefits of using BFA to represent SMT for regular expressions, we make the following contributions:
\begin{itemize}
  \item We formalize the emptiness problem for Boolean finite automata as a safety problem for a Boolean model checker (Section~\ref{sec:alternating}).
  \item We extend the formalization to support multiple variables of different lengths and bounded-history transition predicates.  These are useful for compactly encoding regular expression extensions such as anchors and word boundaries (Section~\ref{sec:multi-var}).
  \item We demonstrate the effectiveness of this encoding by implementing Qzy, an SMT solver for the theory of regular languages, and comparing to other state-of-the-art solvers on open source benchmarks (Section~\ref{sec:evaluation}).
\end{itemize}

\section{Preliminaries: Automata and Model Checking}

\newcommand{\automaton}{A}
\newcommand{\bfa}{B}
\newcommand{\transsys}{S}
\newcommand{\syms}{\Sigma}
\newcommand{\symsb}{\syms_B}
\newcommand{\sym}{\sigma}
\newcommand{\states}{Q}
\newcommand{\statesb}{\states_B}
\newcommand{\statesbfa}{{\tilde{Q}}}
\newcommand{\state}{q}
\newcommand{\path}{\pi}
\newcommand{\sympath}{\bar{\sym}}
\newcommand{\statepath}{\bar{\state}}
\newcommand{\init}{I}
\newcommand{\initb}{\init_B}
\newcommand{\initbfa}{\tilde{\init}}
\newcommand{\final}{F}
\newcommand{\finalbfa}{\tilde{F}}
\newcommand{\prop}{P}
\newcommand{\trans}{\Delta}
\newcommand{\transb}{\trans_B}
\newcommand{\transbfa}{\tilde{\trans}}
\newcommand{\lang}{L}
\newcommand{\bool}{\{0,1\}}

\newcommand{\naturals}{\mathbb{N}}

This section gives the basic definitions and examples that will be used throughout the paper.  Furthermore it unifies the terminology of model checking and automata.

A \emph{finite-state transition system} $\transsys = (\syms, \states, \init, \trans)$ is described by a finite set of input symbols $\syms$, a finite set of states $\states$, an initial set of states $\init \subseteq \states$, and a transition relation $\trans \subseteq \states \times \syms \times \states$.  A \emph{path} $\pi = (\statepath, \sympath, n)$ where $\statepath = \state_0, \ldots, \state_n$ and $\sympath = \sym_0, \ldots, \sym_{n-1}$ is a sequence of states and inputs such that $\state_0 \in \init$ and $\forall i \in \naturals. \; 0 \le i < n \rightarrow (\state_i, \sym_i, \state_{i+1}) \in \trans$.  A \emph{safety property} is a set of states $\prop \subseteq \states$; the property is \emph{valid} if  there does not exist any path that ends with $\state_n \in \prop$.  

A \emph{non-deterministic finite-state automaton}  (NFA) $\automaton = (\syms, \states, \init, \final, \trans)$ is a finite-state transition system with a fixed safety property $\final$ denoting the \emph{final} or \emph{accepting} states.  A finite-state automaton is said to \emph{accept} a sequence of inputs $\sympath$ if there exists a path $\path = (\statepath,\sympath,n)$ such that $\statepath$ ends in a $\state_n \in \final$.  The \emph{language} $\lang$ accepted by the automaton is the set of all accepted $\sympath$.  By definition, a language is empty if and only if the safety property $\final$ is valid.

NFA are drawn using the standard notation where the states labeled with ``start'' are in $\init$, the double circle states are in $\final$, the edges labeled with elements from $\syms$ are in $\trans$.  Examples are shown in Figures~\ref{fig:automaton-aa} and~\ref{fig:automaton-aaa}.

\begin{figure}
  \centering
  \begin{minipage}[t]{0.47\textwidth}
    \centering
    \begin{tikzpicture}[>=stealth', shorten >=1pt, auto, node distance=2 cm, scale = 1, transform shape]
\node[initial, state, accepting] (n1) {$1$};
\node[state]                     (n2) [right of=n1] {$2$};
\path[->]
  (n1) edge [below]             node [align=center] {$\texttt{a}$} (n2)
  (n2) edge [bend right, above] node [align=center] {$\texttt{a}$} (n1);
\end{tikzpicture}%
    \caption{Automaton $\automaton_1$ language $(\texttt{a} \cdot \texttt{a})^*$}
    \label{fig:automaton-aa}
  \end{minipage}\quad
  \begin{minipage}[t]{0.5\textwidth}
    \centering
    \begin{tikzpicture}[>=stealth', shorten >=1pt, auto, node distance=2 cm, scale = 1, transform shape]
\node[initial, state, accepting] (n3) {$3$};
\node[state]                     (n4) [right of=n3] {$4$};
\node[state]                     (n5) [right of=n4] {$5$};
\path[->]
  (n3) edge [below]             node [align=center] {$\texttt{a}$} (n4)
  (n4) edge [below]             node [align=center] {$\texttt{a}$} (n5)
  (n5) edge [bend right, above] node [align=center] {$\texttt{a}$} (n3);
\end{tikzpicture}%
    \caption{Automaton $\automaton_2$ language $(\texttt{a} \cdot \texttt{a} \cdot \texttt{a})^*$}
    \label{fig:automaton-aaa}
  \end{minipage}
\end{figure}

A \emph{hardware model checker} is a decision procedure that determines whether or not a safety property of a finite-state transition system is valid.  In a hardware model checker, the set of symbols $\syms$ and the set of states $\states$ are represented as Boolean vectors $\symsb = \bool^k$ and $\statesb = \bool^m$ equipped with bit-level accessors, $\symsb^i:\symsb\rightarrow \bool$ for $0\leq i < k$ and $\statesb^i:\statesb\rightarrow \bool$ for $0\leq i < m$. The values $k$ (number of symbol bits) and $m$ (number of state bits) are the minimal values such that $2^k \ge |\syms|$ and $2^m \ge |\states|$.

The initial state set $\init$ is represented as a Boolean function $\initb : \statesb \rightarrow \bool$.  Similarly, properties $\prop$ (and $\final$ in NFAs) are Boolean functions on $\statesb$.  The transition relation is represented as a Boolean function $\transb : \statesb \times \syms \times \statesb \rightarrow \bool$.  When representing such a function as a Boolean expression, next-state variables, indices of $\statesb$, are distinguished from current state variables, indices of \textit{the other} $\statesb$, with a prime. E.g.  $\trans = \{(10,11,01)\}$ can be represented by $\transb = \statesb^1 \wedge \neg \statesb^0   \wedge   \Sigma^1\wedge\Sigma^0 \wedge \neg {\statesb^1}' \wedge {\statesb^0}'$.

If a finite-state transition system is invalid, the model checker produces a \emph{counterexample}, a path that ends with a state for which $\prop$ evaluates to $1$.  For an NFA, a counter example to its accepting property is (a path corresponding to) an accepted word of its language.

Satisfiability modulo the theory of regular language membership is described by the following language:
\begin{align*}
  \mathrm{SMT}_\mathrm{RL} ::= & \ \mathrm{SMT}_\mathrm{RL} \wedge \mathrm{SMT}_\mathrm{RL} \\
   | &\ \mathrm{SMT}_\mathrm{RL} \vee \mathrm{SMT}_\mathrm{RL}\\
   | &\ \neg \mathrm{SMT}_\mathrm{RL}\\
   | &\ x \in \lang
\end{align*}
where $x$ is any variable and $L$ is the accepted language of any regular expression.

The semantics of this language are the obvious Boolean logic semantics combined with language membership as described above. We consider two variants of this language.  In the first variant, it is assumed that there is a single variable.  This variant is consistent with existing automata-based approaches and is addressed in Section~\ref{sec:alternating}.  The second variant drops the single variable constraint and will be addressed in Section~\ref{sec:multi-var}.

A \emph{Boolean finite automaton} (BFA) $\bfa = (\syms, \statesbfa, \initbfa, \finalbfa, \transbfa)$ is a finite-state transition system composed of a final assignment, $\finalbfa:\statesbfa\rightarrow \bool$, and Boolean functions, $\initbfa : 2^\statesbfa \rightarrow \bool$ and $\transbfa : \statesbfa \times \syms \rightarrow (2^\statesbfa \rightarrow \bool)$. We will equivocate elements $\state\in \statesbfa$ with the corresponding bit-level accessor, $\state:2^\statesbfa\rightarrow \bool$, so that subsets of $2^\statesbfa$ are naturally represented as Boolean expressions with literals in $\statesbfa$.

To interpret a BFA as an NFA $(\syms, \states, \init, \final, \trans)$, put $Q=2^\statesbfa$. $\init$ is the set indicated by $\initbfa$.  $\final$ is the singleton indicated by $\left(\bigwedge_{\finalbfa(q)=1}\  q \right) \wedge \left( \bigwedge_{\finalbfa(q)=0} \neg q \right)$. Finally, transitions are defined as follows
\begin{align*}
  \trans = \Set{(y,\sym,z) \in \states \times \syms \times \states |  \forall\state\in \statesbfa .\ \state(z) = \transbfa(q,\sym)(y) }
\end{align*}

The resulting NFA acceptance corresponds to a series of substitutions in Boolean expressions as follows.  Begin with $\initbfa$.  A transition on an input symbol $\sym$ replaces each state $\state_i$ occurring in the current formula with $\transbfa(\state_i,\sym)$.  A sequence of symbols is accepted iff the formula is true after each constituent $\state_i$ is replaced with $\final(\state_i)$.

An NFA $(\syms, \states, \init, \final, \trans)$ can be converted to a BFA $(\syms, \initbfa, \finalbfa, \transbfa)$ as follows.  Put $\statesbfa=\states$, $\initbfa = \bigvee_{\state\in\init} q$, $\finalbfa$ is the indicator function for $\final$, and
\begin{align*}
  \transbfa(\state, \sym) = \bigvee \{ \state' \ | \  (\state, \sym, \state') \in \trans \}
\end{align*}
where we've made use of the equivocation between $\state\in \states$ and $q:2^\statesbfa\rightarrow \bool$.

\begin{example}[Convert an NFA to a BFA]
  Consider the NFA $\automaton_1$ in Figure~\ref{fig:automaton-aa}.  We have
  \begin{align*}
    \bfa_1: & \syms = \{\texttt{a}\} \quad \quad \statesbfa = \{\state_1, \state_2\} \quad \finalbfa = [\state_1\mapsto 1, \state_2\mapsto 0] \\
            & \quad \initbfa = \state_1  \quad \transbfa(\state_1,\texttt{a}) = \state_2 \quad \transbfa(\state_2, \texttt{a}) = \state_1
  \end{align*}
\end{example}

\begin{example}[Check if a string is accepted by a BFA]
  The NFA $\automaton_1$ accepts all even-length sequences of $\texttt{a}$ symbols as input.  The corresponding BFA $\bfa_1$ does as well.  It therefore accepts the empty string and $\texttt{aa}$, but rejects $\texttt{a}$.  The state of a BFA is a Boolean formula written $G_i$, where $i$ is the step in the execution:
  \begin{align*}
    G_0 = \initbfa = \state_1  \quad G_1 = \state_2 \quad G_2 = \state_1
  \end{align*}
  Note that to step from $G_0$ to $G_1$, each occurrence of $\state_1$ was replaced with $\state_2$ because $\transbfa(\state_1,\texttt{a}) = \state_2$, similarly each occurrence of $\state_2$ was replaced with $\state_1$.  Since the final state is $\state_1$, substituting true for $\state_1$, and false for $\state_2$ in each formula gives $G_0 = \mathrm{true}$, $G_1 = \mathrm{false}$, and $G_2 = \mathrm{true}$, thus confirming that the empty string and the string $\texttt{aa}$ are accepted and the string $\texttt{a}$ is rejected.
\end{example}

BFA can be efficiently and directly combined solely through syntactic manipulations of the initial function.  The manipulations are exactly the Boolean logic equivalent of the corresponding automata operation:
\begin{align*}
  &(\syms_1, \statesbfa_1, \initbfa_1, \finalbfa_1, \transbfa_1) \cup (\syms_2, \statesbfa_2, \initbfa_2, \finalbfa_2, \transbfa_2) = \\
  &\qquad (\syms_1 \cup \syms_2, \statesbfa_1 \uplus \statesbfa_2, \finalbfa_1 \uplus \finalbfa_2, \boxed{\initbfa_1 \vee \initbfa_2}, \transbfa_1 \uplus \transbfa_2) \\
  &(\syms_1, \statesbfa_1, \initbfa_1, \finalbfa_1, \transbfa_1) \cap (\syms_2, \statesbfa_2, \initbfa_2, \finalbfa_2, \transbfa_2) = \\
  &\qquad (\syms_1 \cup \syms_2, \statesbfa_1 \uplus \statesbfa_2, \finalbfa_1 \uplus \finalbfa_2, \boxed{\initbfa_1 \wedge \initbfa_2}, \transbfa_1 \uplus \transbfa_2) \\
  &(\syms, \statesbfa, \initbfa, \finalbfa, \transbfa)^\complement = (\syms, \statesbfa, \boxed{\neg \initbfa}, \finalbfa, \transbfa)
\end{align*}
where ``$\uplus$'' indicates disjoint union on sets and coproduct on functions, i.e. for $i=1,2$ and functions $f_i:X_i\rightarrow Y$ we define $f_1\uplus f_2:X_1\uplus X_2\rightarrow Y$ by $f_1\uplus f_2 (x)=f_i(x)$ for $x\in X_i$.

\begin{example}[BFA combination] 
  \label{ex:bfa-combination}
  The combination $\bfa_3 = \bfa_1^\complement \cup \bfa_2$ (corresponding to the structure of \lstinline|student_discount| being false), where $\bfa_1$ and $\bfa_2$ are derived from $\automaton_1$ and $\automaton_2$ in Figures~\ref{fig:automaton-aa} and~\ref{fig:automaton-aaa} respectively is:
  \begin{align*}
    &\bfa_1 = (\{\texttt{a}\},\{\state_1,\state_2\}, \state_1 , [\state_1\mapsto 1, \state_2\mapsto 0], [(\state_1,\texttt{a}) \mapsto \state_2, (\state_2,\texttt{a}) \mapsto \state_1]) \\
    &\bfa_2 = (\{\texttt{a}\},\{\state_3,\state_4,\state_5\}, \state_3 , [\state_3\mapsto 1, \{\state_4,\state_5\}\mapsto 0], [(\state_3,\texttt{a}) \mapsto \state_4, (\state_4,\texttt{a}) \mapsto \state_5, (\state_5,\texttt{a}) \mapsto \state_3 ]) \\
    &\bfa_3 = (\{\texttt{a}\},\{\state_1,\state_2,\state_3,\state_4,\state_5\}, \neg \state_1 \vee \state_3 , [\{\state_1,\state_3\}\mapsto 1, \{\state_2,\state_4,\state_5\}\mapsto 0], \transbfa_3) \\
    &\qquad \textrm{where } \transbfa_3 = [(\state_1,\texttt{a}) \mapsto \state_2, (\state_2,\texttt{a}) \mapsto \state_1, (\state_3,\texttt{a}) \mapsto \state_4, (\state_4,\texttt{a}) \mapsto \state_5, (\state_5,\texttt{a}) \mapsto \state_3 ]
  \end{align*}
\end{example}

\section{Model Checking a Boolean Finite Automaton}
\label{sec:alternating}

While the emptiness (or universality) of a BFA can be determined by a depth-first search, a more efficient path to answering this question is to convert the BFA into a transition system suitable for solving with a hardware model checker.  By itself, this is nearly sufficient to solve satisfiability problems modulo regular language membership.  If the regular languages $L$ are represented as regular expressions $R$, the following partial transformation reduces any problem involving a single variable $x$ to a single BFA:
\begin{align*}
  \mathrm{SMT}[x \in B_1 \wedge x \in B_2 \mapsto x \in B_1 \cap B_2] \quad
  \mathrm{SMT}[x \in B_1 \vee x \in B_2 \mapsto x \in B_1 \cup B_2] \\
  \mathrm{SMT}[\neg (x \in B_1) \mapsto x \in B_1^\complement] \quad
  \mathrm{SMT}[x \in L \mapsto x \in \textrm{NFA-to-BFA} \circ \textrm{Re-to-NFA}(L)]
\end{align*}
where Re-to-NFA performs the standard Thompson encoding~\cite{thompson:1968} and NFA-to-BFA performs the translation from NFA into BFA given in the previous section.  The application of every possible rewriting we call the application of SMT-to-BFA.

\begin{example}[SMT to BFA]
  The formula $\neg (x \in (\texttt{a} \cdot \texttt{a})^*) \vee (x \in (\texttt{a} \cdot \texttt{a} \cdot \texttt{a})^*)$ splits into two regular language membership queries as shown in Figures~\ref{fig:automaton-aa} and~\ref{fig:automaton-aaa}.  Those NFA are produced by the Thompson encoding/NFA-to-BFA operation.  The resulting BFA from the NFA-to-BFA operations are the same as $\bfa_1$ and $\bfa_2$ from Example~\ref{ex:bfa-combination}.  The resulting problem is rewritten in the following stages:
  \begin{align*}
    \neg (x \in \bfa_1) \vee (x \in \bfa_2) \quad = \quad x \in \bfa_1^\complement \cup \bfa_2 \quad = \quad x \in \bfa_3
  \end{align*}
\end{example}

The resulting BFA $(\syms, \statesbfa, \initbfa, \finalbfa, \transbfa)$ maps
naturally to the problem format of a hardware model checker, $(\syms, \statesb,
\initb, \transb)$, as follows:

$\states$ is ordered so that $\state_i \in \states$ is translated into a
bit-level $\statesb^i$, and the initial state is simply the variable name replacement, $\initb = \initbfa[\state_i\mapsto \statesb^i]$.

The property tested is the negation of the final condition:
\begin{align*}
  \prop = \left( \bigvee_{q_i\in \finalbfa}\  \neg\statesb^i \right) \vee \left( \bigvee_{q\not\in \finalbfa} \statesb^i \right)
\end{align*}
and finally, the transition relation is constructed to effect the BFA substitution:
\begin{align*}
  \bigwedge_{\state_i \in \states}
  {\statesb^i} = \left( \bigvee_{\sym \in \syms} x = \sym \wedge \transbfa(\state_i, \sym)[\state_j \mapsto {\statesb^j}'  \textrm{ for } \state_j \in \states ] \right) 
\end{align*}
The translation process from a BFA to a transition system is known as BFA-to-TS.

\begin{example}[BFA to Transition System]
  The BFA $\bfa_3$ is translated to the following transition system $(\{\texttt{a}\}, (\statesb^1, \ldots, \statesb^5), \initb, \transb)$ where $\initb = \neg\statesb^1 \vee \statesb^3$ and
  \begin{align*}
    \transb(\statesb', x, \statesb) = &{\statesb^1} = (x = \texttt{a} \wedge {\statesb^2}') \wedge {\statesb^2} = (x = \texttt{a} \wedge {\statesb^1}') \wedge {\statesb^3} = (x = \texttt{a} \wedge {\statesb^4}') \\
    \wedge &{\statesb^4} = (x = \texttt{a} \wedge {\statesb^5}')  \wedge {\statesb^5} = (x = \texttt{a} \wedge {\statesb^3}')
  \end{align*}
  The property to be verified is $\neg \statesb^1 \vee \neg \statesb^3 \vee \statesb^2 \vee \statesb^4 \vee \statesb^5$.  Because the transition system is verifying a universal property instead of finding an example of an existential property, the accepting condition has been negated.
\end{example}

\begin{theorem}[Polynomial time encoding]
  For all $\textrm{SMT}_\textrm{RL}$, TS such that $\textrm{TS},P = \textrm{BFA-to-TS} \circ \textrm{SMT-to-BFA} (\textrm{SMT}_\textrm{RL})$, the process takes at most polynomial number of steps and produces a TS that is at most polynomial in the size of the $\textrm{SMT}_\textrm{RL}$ problem.
\end{theorem}

In fact, the encoding time is completely linear with the exception of epsilon elimination that is required after the use of Thompsons encoding.  Because epsilon elimination is at worst the transitive closure of a directed graph, it is bounded by $O(n^3)$, but in practice, it is quite efficient and the encoding time is thus negligible compared to the solve time needed by the model checker.

Because hardware model checkers operate on bits, one bit is built for each state in the BFA.  Similarly symbols must be encoded into bits.  Any bit-wise encoding of a $x \in \syms_c \subseteq \syms$ predicate will work.  This allows for representing whole sets of transitions with a single compact formula.  However, since the number of bits is linear in the number of BFA states, the number of transition system states is exponential in the number of BFA states, thus effectively performing an on-demand, \emph{lazy} determinization of the original NFA.  The SMT to transition system encoding together with a sound and complete model checking algorithm is a decision procedure for satisfiability modulo regular language membership with one variable.

\begin{theorem}[Decision procedure]
  For all symbol sequences $x$, $x \models \textrm{SMT}_\mathrm{RL}$ iff $x$ is a counterexample to the property of the transition system $\textrm{BFA-to-TS} \circ \textrm{SMT-to-BFA} (\textrm{SMT}_\mathrm{RL})$.
\end{theorem}

\section{Satisfiability Modulo Regular Language Inclusion}
\label{sec:multi-var}

To decide the satisfiability of a full Boolean combination of regular language membership predicates, we need to extend the formalism.  The transformation given in the previous section does not handle two key challenges that will be addressed in this section: (1) multiple variables and (2) anchors (the occurrence of beginning-of-string and ending-of-string restrictions in the middle of a regular expression).  It turns out that the handling of multiple variables and anchors are intertwined.

\subsection{Multiple variables}

To address the multi-variable problem we will go back to the NFA representation and address problems there.  The obvious solution is to replace the symbols $\syms$ with a vector of symbols $\syms_1 \times \ldots \times \syms_n$, one for each different variable $x_i$ for $i \in 1 \ldots n$ in the SMT problem.  Of course each NFA is only associated with a single variable, so transitions occur when the current input to $x_i$ matches the appropriate element from $\syms_i$.

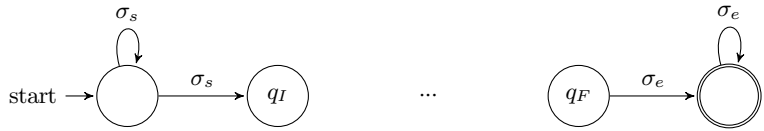
\begin{figure}
  \centering
  \begin{tikzpicture}[>=stealth', shorten >=1pt, auto, node distance=2 cm, scale = 1, transform shape]
\node[initial, state]   (n1) {};
\node[state]            (n2) [right of=n1]{$\state_I$};
\node                   (dots)[right of=n2] {$\cdots$};
\node[state]            (n3) [right of=dots]{$\state_F$};
\node[state, accepting] (n4) [right of=n3] {};
\path[->]
  (n1) edge [loop above] node [align=center] {$\sym_s$} (n1)
  (n1) edge [above]      node [align=center] {$\sym_s$} (n2)
  (n3) edge [above]      node [align=center] {$\sym_e$} (n4)
  (n4) edge [loop above] node [align=center] {$\sym_e$} (n4);
\end{tikzpicture}%
  \caption{Automaton with initial state $\state_I$ and final state $\state_F$ is extended with start symbols $\sym_s$ and end symbols $\sym_e$.}
  \label{fig:nfa-extend}
\end{figure}

While this is sufficient to allow the $\textrm{SMT-to-BFA}$ function to be fully applied, the resulting transition systems are not equivalent to the original SMT problem.  The problem is that this implicitly forces all variables to have the same length.  For example, if the variable $x_1$ matched $\automaton_1$, and $x_2$ matched $\automaton_2$, the shortest non-zero-length satisfying assignment would be that $x_1 = x_2 = \texttt{a} \cdot \texttt{a} \cdot \texttt{a} \cdot \texttt{a} \cdot \texttt{a} \cdot \texttt{a}$, whereas if considered independently, the shortest non-zero-length satisfying assignment is $x_1 = \texttt{a} \cdot \texttt{a}$ and $x_2 = \texttt{a} \cdot \texttt{a} \cdot \texttt{a}$.  To circumvent the problem, we introduce two extra symbols to every NFA: $\sym_s$ is the start of the sequence and $\sym_e$ is the end of the sequence.  Every NFA is extended as shown in Figure~\ref{fig:nfa-extend}.

\begin{example}[Start and end extension]
  If solving the problem $x_1 \in A_1 \wedge x_2 \in A_2$, where $A_1$ and $A_2$ are as defined in Figures~\ref{fig:automaton-aa} and~\ref{fig:automaton-aaa} respectively, the shortest non-zero-length satisfying assignment after start and end extension is:
  \begin{align*}
  x_1 = \sym_s \cdot \texttt{a} \cdot \texttt{a} \cdot \sym_e \cdot \sym_e \qquad x_2 = \sym_s \cdot \texttt{a} \cdot \texttt{a} \cdot \texttt{a} \cdot \sym_e
  \end{align*}
  The two answers are the same length, but are padded with start and end symbols so that the actual satisfying assignment (stripped of start/end symbols) allows different lengths.
\end{example}

With this simple extension, multiple variable problems can be transformed into a single transition system using the $\textrm{BFA-to-TS} \circ \textrm{SMT-to-BFA}$ operation without overly restricting the problem.  This means that all unsatisfiable answers are genuinely unsatisfiable.

A new issue arises, answers reported as satisfiable in the encoding may not be satisfiable in the original problem.  For example, consider the SMT problem: $\neg (x \in \syms^*)$.  There should be no satisfying assignments to $x$.  However under this new model, there are many.  For example, $x = \sym_e \cdot \sym_s$ is a satisfying assignment because invalid sequences of start and end symbols are satisfying assignments to the negation of a regular language membership predicate.  To remedy this problem the SMT problem is transformed with additional constraints:
\begin{align*}
  \textrm{SMT}_\textrm{RL} \wedge \bigwedge_{x \in X} x \in \syms^*
\end{align*}
This enforces that every satisfying assignment must be a \emph{valid} sequence of start, end, and symbols from $\syms$.

The combination of the start and end extension with the augmented SMT problem is known as the MV-Rewrite operation.

\begin{theorem}[Multi-variable decision procedure]
    For any $\textrm{SMT}_\textrm{RL}$, $\textrm{TS}$, and $P$ such that  $\textrm{TS},P = \textrm{BFA-to-TS} \circ \textrm{SMT-to-BFA} \circ \textrm{MV-Rewrite}(\textrm{SMT}_\textrm{RL})$, there exists a model $\bar{x}, \textrm{TS} \models \neg P$ iff there exists a satisfying assignment to $\textrm{SMT}_\textrm{RL}$.
\end{theorem}

\subsection{Anchoring regular expressions}

A common feature in regular expressions is anchors.  In regular expressions these are \texttt{\^} for the beginning of the string and \texttt{\$} for the ending of the string.  They both match an empty sequence of symbols, but \texttt{\^} only matches if there is no preceding symbol and \texttt{\$} only matches if there is no succeeding symbol.  While it is possible and efficient to remove these from regular expressions by forcing all paths preceding (resp. succeeding) a \texttt{\^} (\texttt{\$}) to be empty, this is a non-trivial implementation strategy.  Furthermore, there are other zero-length operators (Section~\ref{sec:practical}) that can, in the worst case, cause a quadratic increase in NFA size to remove.

In this section we generalize the notion of anchors to zero-length, bounded-history predicates and build upon our use of start and end extension to simply handle any anchor used in regular expressions.  This generalization enables a linear increase in the encoding size to represent an exponential (in the length of the history) increase in the problem size.  This general form is then applied to anchors (history size one) and other operators (up to history size two) used in regular expressions.

The generalization expands $\syms$ into $\syms_{-h+1} \times \ldots \times \syms_0$ for a history of length $h$.  Each transition in $\trans$, $\transbfa$, or $\transb$ selects from whichever $\syms_i$ necessary.  For example, to proceed only if a string begins with the symbol \texttt{a} (in other words $\mbox{\texttt{\^}} \cdot \texttt{a}$), the transition would require that $\sym_{-1} = \sym_s$ and $\sym_0 = \texttt{a}$.

The expression of this as an extension of the symbols leads to a simple implementation in the transition system.  An extra $|\symsb|*(k-1)$ bits are added to the state to store historical inputs and then the transition is constrained so that ${\symsb^i}'_j = {\symsb^i}_{j-1}$ to preserve the history.  The initialization of these bits would be problematic if it were not for the introduction of the $\sym_s$ language extension.  This gives an initial value to all of the history that allows transitions that depend on the value before the beginning of the string to succeed.  Furthermore, by adding $\sym_e$ to the end, dependence on the end of the string can match on $\sym_e$ and ensure that the end is reached.

\section{Practicalities}
\label{sec:practical}

Although the BFA formalism leads to a problem readily solved with model checking, solving the \textit{reverse} automaton is preferred for two reasons.  The first reason is that the final set for a BFA is a singleton, so it is much better to start with it rather than the BFA initial set, which is arbitrary.  The second observation is that reversing the BFA has the effect of making the transition equations \textit{deterministic} -- next states can be written as a function of previous states.  Having reversed the language recognized by the BFA in BFA-to-TS, we reverse the order of the constituent NFAs during NFA-to-BFA so that solutions are overall correct.

\begin{example}[Practical anchor handling]
  Consider the regular expression $(\texttt{a}|(\mbox{\texttt{\^}}\cdot \texttt{b})|(\texttt{c} \cdot \mbox{\texttt{\$}}))^*$.  The length three symbol sequences this matches are $\texttt{a} \cdot \texttt{a} \cdot \texttt{a}$, $\texttt{b} \cdot \texttt{a} \cdot \texttt{a}$, $\texttt{a} \cdot \texttt{a} \cdot \texttt{c}$, and $\texttt{b} \cdot \texttt{a} \cdot \texttt{c}$.  Note the absence of $\texttt{c} \cdot \texttt{a} \cdot \texttt{b}$ due to the restrictions of the anchors.
  
  The resulting (augmented) NFA is shown in Figure~\ref{fig:automaton-epsilon}.  Because anchors match a sequence of length zero, this is equivalent to an NFA with epsilons except that the epsilons have a history dependence.  After epsilon elimination (Figure~\ref{fig:automaton-anchored}) the anchors all precede an actual symbol.  Of course, some of the transitions are impossible and can be eliminated, but the history dependence remains -- in particular for the \texttt{\^} anchor.
  
  The resulting transition is shown below.  According to the optimization mentioned above, the NFA is reversed and the transition system is also reversed resulting in a transition function rather than a transition relation.
  \begin{align*}
  & P = \neg \statesb^5 \qquad \initb = \statesb^1 \wedge \neg \statesb^2 \wedge \neg \statesb^4 \wedge \neg \statesb^5 \wedge \statesb^\syms = \sym_s \\
    &\transb = {\statesb^1}' = \left(x = \sym_s \wedge \statesb^1 \right)
               \wedge {\statesb^2}' = \left(x = \texttt{a} \wedge \statesb^2 \right)
                                  \vee \left(x = \texttt{b} \wedge \statesb^{\syms - 1} = \sym_s \wedge \statesb^2 \right) \\
            &{} \wedge {\statesb^4}' = \left(x = \texttt{c} \wedge \statesb^2 \right)
               \wedge {\statesb^5}' = \left(x = \sym_e \wedge \statesb^5 \right)
                                  \vee \left(x = \sym_e \wedge \statesb^4 \right)
                                  \vee \left(x = \sym_e \wedge \statesb^2 \right)
               \wedge {\statesb^\syms}' = x
  \end{align*}
\end{example}

\begin{figure}[t]
    \centering
    \begin{minipage}[t]{0.47\textwidth}
        \centering
        \begin{tikzpicture}[>=stealth', shorten >=1pt, auto, node distance=1.75 cm, scale = 1, transform shape]
\node[initial, state]   (n1) {1};
\node[state]            (n2) [right of=n1]{2};
\node[state]            (n3) [xshift=-0.5cm, above of=n2]{3};
\node[state]            (n4) [right of=n2] {4};
\node[state, accepting] (n5) [above of=n4] {5};
\path[->]
  (n1) edge [loop above] node [align=center] {$\sym_s$} (n1)
  (n1) edge [above]      node [align=center] {$\sym_s$} (n2)
  (n2) edge [loop below] node [align=center] {$\texttt{a}$} (n2)
  (n2) edge [bend left, below]  node [align=center] {$\texttt{c}$} (n4)
  (n2) edge [bend left, left]   node [align=center] {$\mbox{\texttt{\^}}$} (n3)
  (n2) edge [below right]        node [align=center] {$\sym_e$} (n5)
  (n3) edge [bend left, left]   node [align=center] {$\texttt{b}$} (n2)
  (n4) edge [bend left, below]  node [align=center] {$\texttt{\$}$} (n2)
  (n5) edge [loop left]         node [align=center] {$\sym_e$} (n5);
\end{tikzpicture}%
        \caption{History dependence w/epsilons}
        \label{fig:automaton-epsilon}
    \end{minipage}\quad
    \begin{minipage}[t]{0.5\textwidth}
        \centering
        \begin{tikzpicture}[>=stealth', shorten >=1pt, auto, node distance=1.75 cm, scale = 1, transform shape]
\node[initial, state]   (n1) {1};
\node[state]            (n2) [right of=n1]{2};
\node[state]            (n4) [right of=n2] {4};
\node[state, accepting] (n5) [above of=n4] {5};
\path[->]
  (n1) edge [loop above] node [align=center] {$\sym_s$} (n1)
  (n1) edge [above]      node [align=center] {$\sym_s$} (n2)
  (n2) edge [loop below] node [align=center] {$\texttt{a}$} (n2)
  (n2) edge [bend left, below]  node [align=center] {$\texttt{c}$} (n4)
  (n2) edge [loop above]   node [align=center] {$\mbox{\texttt{\^}} \cdot \texttt{b}$} (n2)
  (n2) edge [below right]        node [align=center] {$\sym_e$} (n5)
  (n4) edge [below]  node [align=center] {$\texttt{\$} \cdot \texttt{a}$} (n2)
  (n4) edge [bend left, below]  node [align=center] {$\texttt{\$} \cdot \mbox{\texttt{\^}} \cdot \texttt{b}$} (n2)
  (n4) edge [right]  node [align=center] {$\texttt{\$} \cdot \sym_e$} (n5)
  (n5) edge [loop left]         node [align=center] {$\sym_e$} (n5);
\end{tikzpicture}%
        \caption{History dependence w/o epsilons}
        \label{fig:automaton-anchored}
    \end{minipage}
\end{figure}

To be able to determine the satisfiability of problems involving regular expressions written by software developers, efficiency can be gained by exploiting common idioms.  A common idiom in Perl-compatible regular expressions (PCRE) is the use of character classes.  Character classes allow the expression of ranges of characters such as $[\texttt{a}-\texttt{z}]$.  When deriving a transition system, rather than creating a series of input predicates $x = \texttt{a}$, $x = \texttt{b}$, etc., the range can be compressed using bitvector arithmetic techniques.  For instance, the character class $[\texttt{a}-\texttt{z}]$ is represented as the bitvector predicate $97 \le_8 x \le_8 122$, where the $8$ signifies the number of bits used in the comparison.

Another practical problem is the use of international characters.  Since regular expressions are used throughout the world, it is expected that characters are selected from the thousands that are part of the Unicode specification.  We handle this through the use of the UTF-8 encoding.  This multi-byte encoding is equivalent to ASCII for characters in the range 0 to 127.  Beyond that, multiple bytes are required.  In the implementation we support only single byte input ranges and thus the regular expressions are modified to encode all of the multi-byte possibilities.  This means that the regular expressions that are input may be significantly less complicated (in appearance) than the regular expression that is presented to the algorithm.

Finally, beginning-of-string and ending-of-string anchors are not the only form of anchor used in regular expressions.  Also common are beginning-of-line, ending-of-line, word-boundary, and not-word-boundary anchors.  Given the above encoding, these anchors are easily expressible as predicates that look at either the previous symbol, the next symbol, or both.  Word boundary anchors are perhaps the most problematic because they can cause a significant expansion in the size of the regular expression if removed a priori.  Because of the ability to look a bounded number of symbols into the past, word boundaries present no additional up-front encoding cost in terms of size or complexity.

\subsection{Implementation}

We implemented the complete encoding in a new SMT solver called Qzy.  It is implemented as a C++ library and exposes a programmatic interface similar to that of other SMT solvers.  It uses a sequence of rewrites to achieve the above algorithms.

First, each regular expression is replaced by an NFA with epsilons.  This rewrite is performed by the Re2 library~\cite{re2}.  Re2 is a high-performance regular expression library that internally uses NFA.  It uses an adapted version of Thompsons encoding to produce NFA after having removed all of the Unicode.  Resulting NFA match byte ranges and may have edges that consist only of anchors and other zero-length predicates.  The use of Re2 enables Qzy to support a significant subset of PCRE (only excluding elements that allow PCRE to match non-regular languages).

Qzy then performs epsilon elimination followed by the anchor-aware $\textrm{BFA-to-TS} \circ \textrm{SMT-to-BFA} \circ \textrm{MV-Rewrite}$ algorithm described above.  This results in a compact hardware model checking problem that can be given to any off-the-shelf hardware model checker.

While Qzy can support any AIGER-compatible~\cite{aiger} model checker, by default it uses IC3Ref~\cite{ic3ref}.  In developing Qzy we evaluated a variety of other model checkers including aigbmc~\cite{aiger}, abc~\cite{abc}, IIMC~\cite{iimc}, and NuSMV~\cite{nusmv}.  We found that only IC3Ref and aigbmc (and the bounded model checking engines of IIMC and ABC) offered acceptable performance and of those only IC3Ref produces proofs as well as counterexamples.  We hypothesize that the simplicity of IC3Ref is its key to success.  Because we control the problem generation, we minimize the amount of redundancy that can be eliminated using the preprocessing provided by more advanced model checkers such as abc and IIMC.  By using IC3Ref more time is spent solving the problem and less time is spent preparing the problem to be solved.  In our results we only present IC3Ref.

\section{Evaluation}
\label{sec:evaluation}

In this section we compare the performance of our approach with other solvers for strings and regular expressions. While our algorithm decides the satisfiability of an arbitrary Boolean combination of regular language membership constraints where string variables are unbounded in length, other solvers have adopted different problem domains.  The only directly comparable solver is Norn~\cite{norn}.  Consequently, we restrict the problem domain to a conjunction of possibly negated regular language membership predicates over a single string variable.  This allows comparison against the BRICS automaton library~\cite{brics} and the DPRLE solver~\cite{dprle} in addition to Norn.  We excluded StrSolve~\cite{strsolve}, which is a lazy version of DPRLE due to the unavailability of the code.  Solvers such as CVC4~\cite{cvc4} and Z3-str2~\cite{z3str2} were excluded due to their lack of support for negation of regular language membership and the Hampi solver~\cite{hampi} was excluded due to its lack of support for unbounded strings.

All tests were run on a 3.4GHz Intel processor with 8GB of RAM.  Tests running in Norn, which runs on the Java virtual machine (JVM), were limited to 4GB of heap and had timing measured internally after JVM startup.  Time on all benchmarks was limited to 600s.  Any resource exhaustion is shown as a timeout and thus given a time of 600s, though aside from Norn resources were rarely exhausted. Qzy takes regular expressions with anchors directly, whereas Norn takes regular expressions without anchors.  Anchors were removed for Norn by an untimed separate preprocessing phase. DPRLE and BRICS Automaton take NFA with epsilons as input.  The regular expressions for Norn were transformed into NFA using Thompsons encoding.

We conduct four tests on real-world benchmarks.  The benchmarks come from RegExLib~\cite{regexlib}, a collection of user-submitted regular expressions that match URLs, email addresses, HTML tags, and so forth.  We selected two sets of benchmarks.  The first set (results shown in the first row of Figure~\ref{fig:regexlib}) is pairwise combinations of the 25 regular expressions with the largest syntax trees.  The second set (results shown in the second row of Figure~\ref{fig:regexlib}) is pairwise combinations of the regular expressions that were difficult to determinize using the BRICS Automaton package.

To understand the performance consequences of increasingly large, hard problems, we built four SMT problems that contain difficult to determize regular expressions whose size depends on a parameter $n$.  Note that in a regular expression $\{n\}$ denotes a concatentation of the preceding element $n$ times.  The problems are: satisfiable difference ($x \in \mbox{\texttt{\^}} \cdot [\texttt{01}]^* \cdot \texttt{1} \cdot [\texttt{01}]\{n\} \cdot \texttt{\$} \wedge  x \notin \mbox{\texttt{\^}} \cdot [\texttt{01}]* \cdot \texttt{0} \cdot [\texttt{01}]\{n-1\} \cdot \texttt{\$}$), unsatisfiable difference ($x \in \mbox{\texttt{\^}} \cdot [\texttt{01}]^* \cdot \texttt{1} \cdot \texttt{1} \cdot [\texttt{01}]\{n\} \cdot \texttt{\$} \wedge  x \notin \mbox{\texttt{\^}} \cdot [\texttt{01}]* \cdot \texttt{1} \cdot [\texttt{01}]\{n+1\} \cdot \texttt{\$}$), satisfiable intersection ($x \in \mbox{\texttt{\^}} \cdot [\texttt{01}]^* \cdot \texttt{1} \cdot [\texttt{01}]\{n\} \cdot \texttt{\$} \wedge  x \in \mbox{\texttt{\^}} \cdot [\texttt{01}]* \cdot \texttt{0} \cdot [\texttt{01}]\{n-1\} \cdot \texttt{\$}$), and unsatisfiable intersection ($x \in \mbox{\texttt{\^}} \cdot [\texttt{01}]^* \cdot \texttt{1} \cdot [\texttt{01}]\{n\} \cdot \texttt{\$} \wedge  x \in \mbox{\texttt{\^}} \cdot [\texttt{01}]* \cdot \texttt{0} \cdot [\texttt{01}]\{n\} \cdot \texttt{\$}$).  Performance on these problems gives insight into how different solvers scale on difficult problems.

\subsection{Results}

\begin{figure}
  \centering
  \begin{minipage}{0.32\textwidth}
    \includegraphics[width=\textwidth]{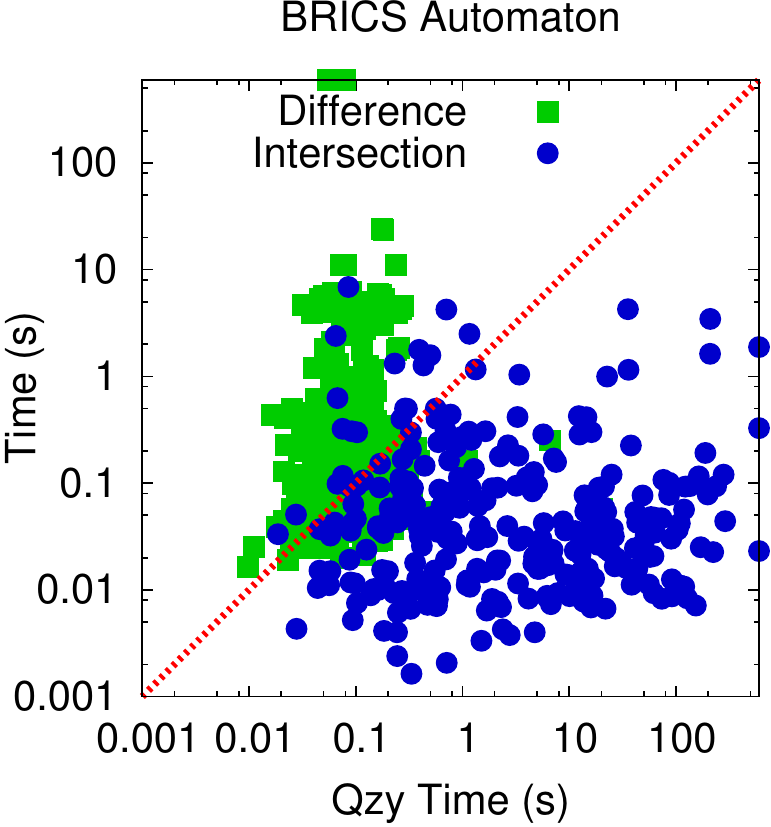}
  \end{minipage}
  \begin{minipage}{0.32\textwidth}
    \includegraphics[width=\textwidth]{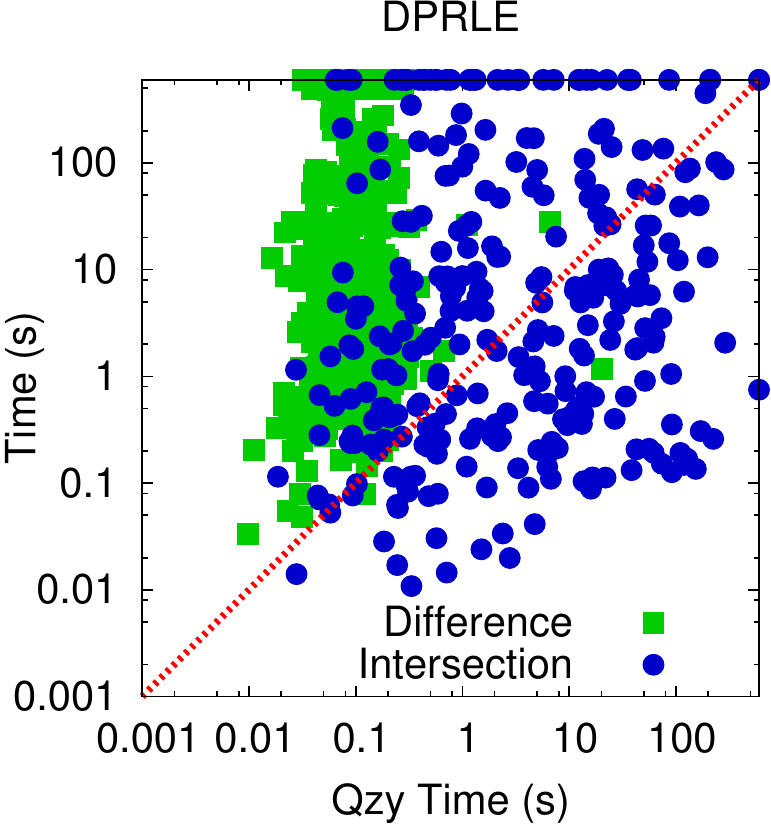}
  \end{minipage}
  \begin{minipage}{0.32\textwidth}
    \includegraphics[width=\textwidth]{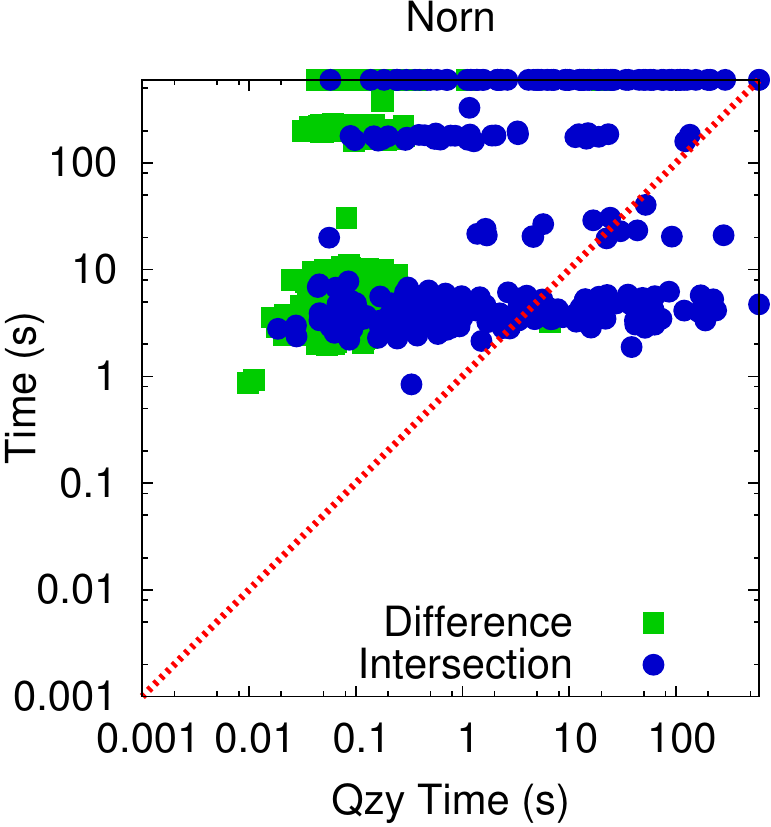}
  \end{minipage} \\[0.5cm]
  \begin{minipage}{0.32\textwidth}
    \includegraphics[width=\textwidth]{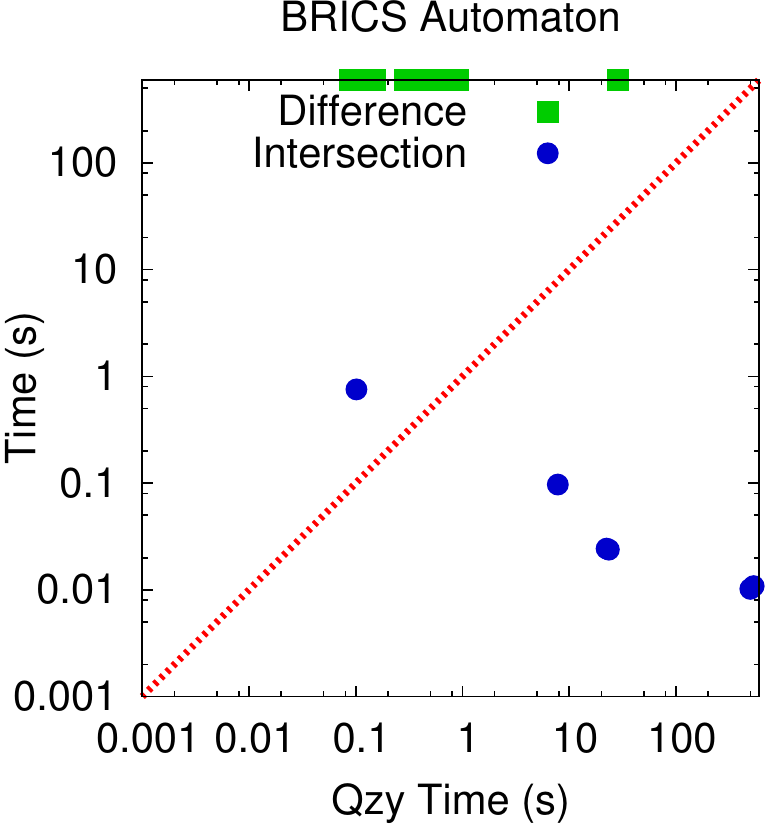}
  \end{minipage}
  \begin{minipage}{0.32\textwidth}
    \includegraphics[width=\textwidth]{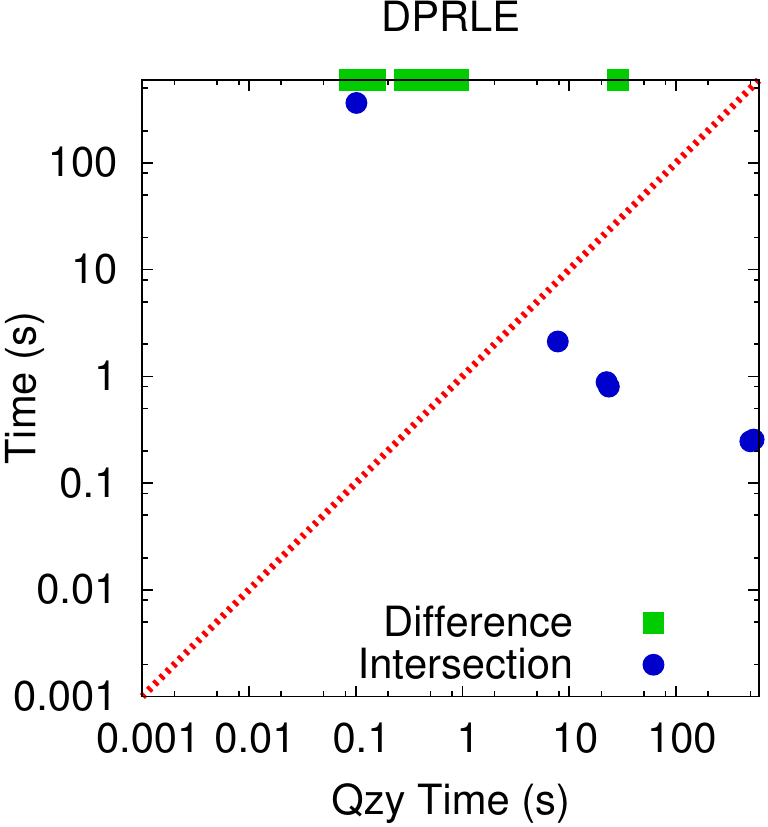}
  \end{minipage}
  \begin{minipage}{0.32\textwidth}
    \includegraphics[width=\textwidth]{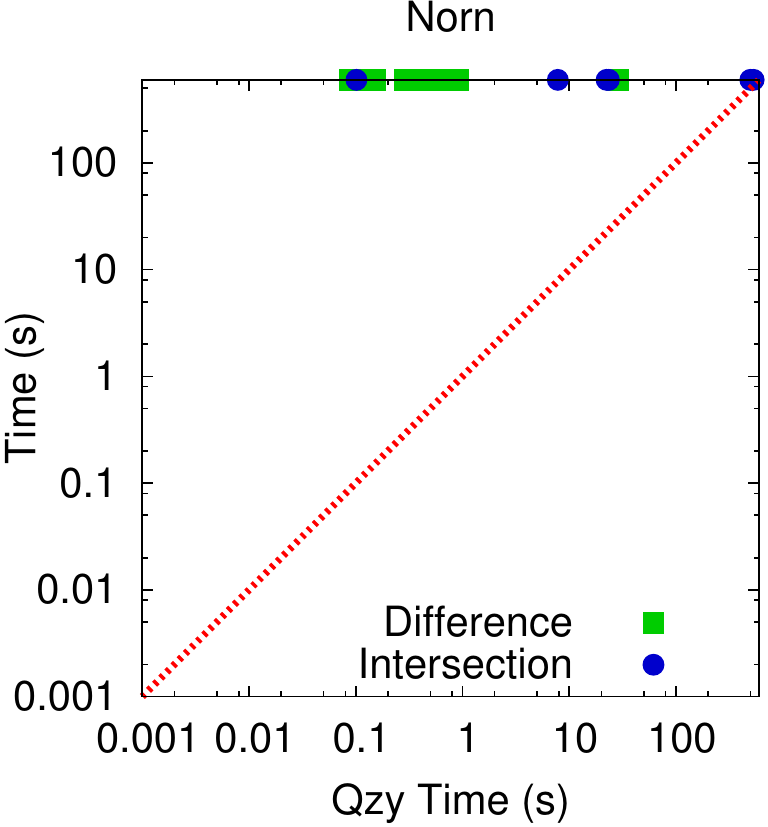}
  \end{minipage}
  \caption{Time to solve problems from RegExLib.  First row are syntactcially large problems.  Second row are difficult to determinize.  Intersection problems are $x \in L_1 \wedge x \in L2$.  Difference problems are $x \in L_1 \wedge x \notin L_2$.}
  \label{fig:regexlib}
\end{figure}

In Figure~\ref{fig:regexlib} we see a comparison of the performance of Qzy against the three other solvers.  If Qzy is faster the dot appears above the line.  The first row (large regular expressions) suggests that large is not correlated with difficult.  The BRICS Automaton package is able to compute the intersection and difference of most pairs with minimal difficulty.  By comparison, Qzy is faster at the harder problem of difference and slower at the easier problem of intersection.  This can largely be explained by the fact that Qzy is a lazy approach.  It is doing more work to manage the laziness than is needed to solve the problem by simply computing the intersection automaton.

The other solvers also suffer from being more clever than BRICS.  DPRLE is slower than Qzy at intersection (it suffers many time outs) and much slower at difference.  Norn is just slow when it comes to regular expressions.  It fails to solve many easy problems in a reasonable amount of time.  Comparatively Qzy is much faster than both, but especially for difference.  This suggests that Qzy is particularly fast when it comes to handling complementation.

The second row of Figure~\ref{fig:regexlib} shows what happens when difficult-to-determinize regular expressions are used.  We see that BRICS and DPRLE perform respectably (often beating Qzy) for the intersection cases.  However, Qzy never suffered a resource exhaustion on any of the hard problems, whereas all competing tools timed out on the hard difference problems.  This further supports Qzy's approach to solving problems.  While it may be slower than other approaches on easier problems, it significantly outperforms other approaches on hard problems.

\begin{figure}
  \centering
  \begin{tabular}{@{$\qquad\quad$}c@{$\quad\qquad\qquad\qquad\qquad\qquad\qquad$} c}
      Difference & Intersection
  \end{tabular}\\
  \includegraphics[width=\textwidth]{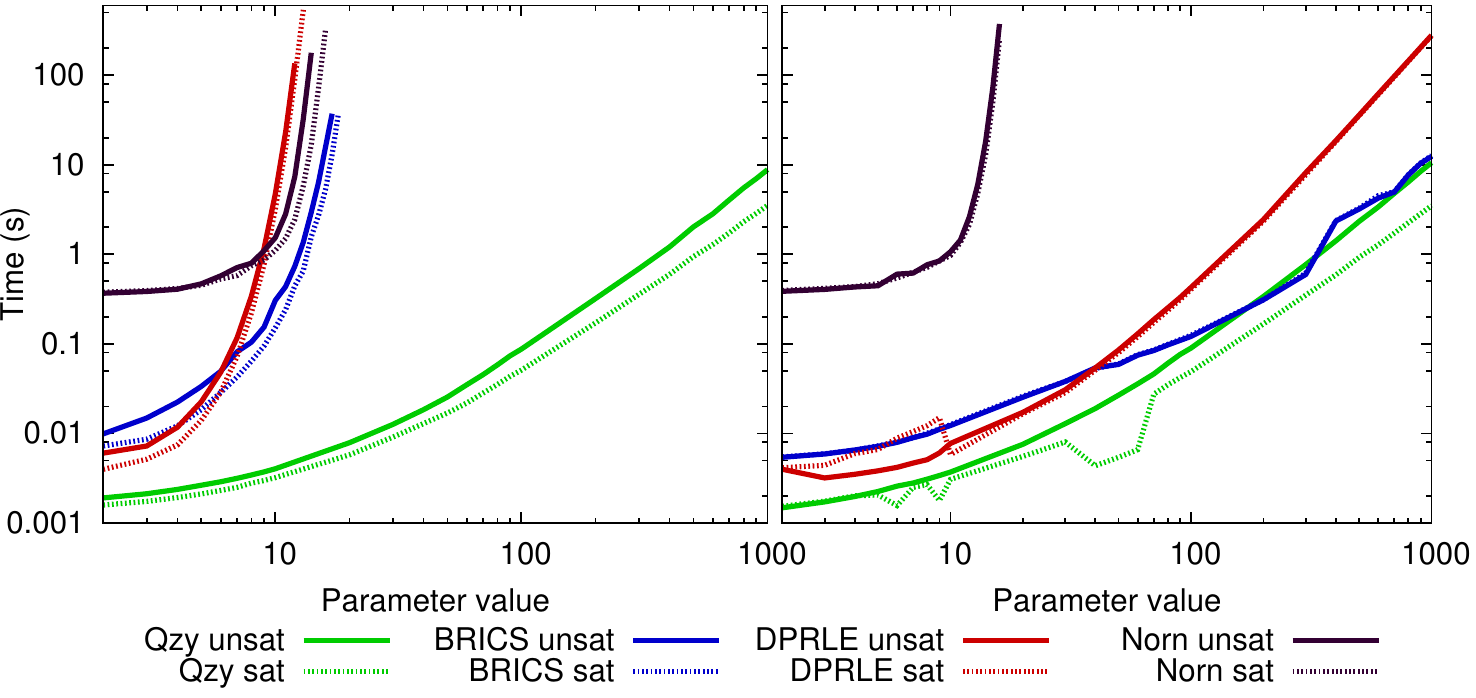}
  \caption{Time to solve problems involving two parametric regular expressions.  The complexity of the problem is determined by a single parameter.}
  \label{fig:synthetic}
\end{figure}

To understand these tradeoffs between easy and hard problems further, we turn to the parametric SMT benchmarks shown in Figure~\ref{fig:synthetic}.  Here we can study the asymptotic behavior by gradually increasing the problem complexity.  We see that for intersection, Qzy is comparable to BRICS Automaton.  The point at which the lines meet, they both appear to have the same asymptotic behavior.  By comparison, DPRLE is much slower.  It seems to scale roughly cubically versus the quadratic behavior of BRICS and Qzy.  We can understand Norn's poor performance on real-world benchmarks by seeing that it simply scales poorly in the complexity of the regular expression.

The parametric difference benchmarks (left of Figure~\ref{fig:synthetic}) show that Qzy is unmatched for complementation.  Due to the generalization techniques that IC3 uses, it maintains a near quadratic behavior on a problem that in the worst case is exponential.  We see the exponential behavior in all of the other solvers.  This suggests that Qzy does best when handing difficult benchmarks.  This further explains the results that we see in the second row of Figure~\ref{fig:regexlib} where Qzy is the only solver to handle any of the difficult difference problems.

We attribute Qzy's performance to the combination of a good encoding with the IC3 algorithm.  Because Qzy produces clean model checking problems we are able to eliminate much of the overhead of hardware model checkers doing preprocessing.  This allows Qzy to be fast at easy problems without becoming unbearably slow at hard problems.

\section{Related Work}

There are a variety of libraries optimized to support automata operations.  DPRLE~\cite{dprle} is a decision procedure for the conjunction of string concatenations included in a regular language or its complement.  This was improved upon for StrSolve~\cite{strsolve} where they made the intersection and complementation operations lazy so as not to incur an upfront exponential cost.  Of course even the lazy version is exponential in the worst case and is heavily dependent on heuristics in the lazy operations.  Our work similarly performs operations lazily, but it does so over an entire SMT problem involving multiple variables, rather than just a conjunction.  Furthermore, because we use IC3, we can benefit from any generalization built-in to IC3.  Unfortunately the authors of StrSolve did not reply to requests for their tool to perform a comparison.

Symbolic automata~\cite{symbautomata} are a technique for more compactly representing automata.  They do not provide any gain in terms of the number of states in a NFA, but they do provide improvements in the representation of edges.  By representing transitions as predicates rather than symbols from an alphabet, they can compactly represent all transitions between the same two states.  It has been shown that many algorithms that can be applied to automata can be directly applied to symbolic automata.  We use a generalization of symbolic automata in our implementation so that we do not have to keep track of each symbol individually.  The generalization is that we support predicates not just on edges that accept a symbol, but also on epsilon edges.  This allows us to represent any combination of bounded history predicates with a single and-inverter graph.

The idea of answering regular language universality queries using model checking originated with Tabakov and Vardi~\cite{tabakov:lpar:2005}.  Wang et al.~\cite{wang:cav:2016} use a similar technique to determine emptiness of a single regular expression that includes intersection.  Wang et al. also use IC3, but neither Tabakov or Wang support multiple variables, complementation, and the full regular subset of PCRE.

It has been shown that anti-chain methods~\cite{antichain} can be effective for solving model checking problems through the lens of automata.  These methods are incomparable to the methods presented here.  They use a different strategy for theorem proving.  It would be interesting future work to compare anti-chain-based model checking to SAT-based model checking on this class of problems.

Finally, there are a variety of SMT solvers for strings.  Hampi~\cite{hampi} is an SMT solver for fixed-length strings with regular expressions and context-free grammars.  Norn~\cite{norn} is an SMT solver for word equations that also supports basic regular expressions.  CVC4~\cite{cvc4} is a general purpose SMT solver that now has a theory for word equations.  It also has basic support for regular expressions, but does not support complementation in any usable fashion.  With more engineering it could support complementation as regular expressions are closed under complementation.  Z3-Str2~\cite{z3str2} is similar to CVC4's string solver except that it operates on top of the Z3~\cite{z3} SMT solver.  All of these have a different goal to the approach presented here.  The focus of these solvers is largely on word equations.  It may be possible that our model-checking-based approach could be added to existing solvers to give them an edge in solving the regular expression parts of their problems.  This is a possible avenue for future work.

\section{Conclusions and Future Work}

We have shown a different way of viewing the problem of SMT solving for regular expressions.  Rather than directly using non-deterministic finite automata, we use Boolean finite automata, which are efficiently intersected and complemented.  We have shown that there is a direct translation from Boolean finite automata to transition systems solvable with a hardware model checker.  In addition, we give a translation that supports the full regular subset of Perl-compatible regular expressions, including word boundaries, anchors, Unicode, and case-insensitivity; and our translation supports multiple independent variables simultaneously in a single problem.  Finally we have demonstrated that the techniques work in practice on a variety of benchmarks collected from a database of regular expressions.

There are several questions that remain open.  Is there a good way to solve word equations along with the model-checking-based regular language membership checks?  Related work includes this functionality, but its adaptation to model checking is not so direct.  Is there an efficient way to incrementally solve these problems with changing goals?  While there has been some effort~\cite{incic3} to make incremental model checkers, this field is young, especially in academia.  We hope to explore these questions in the future expanding upon the core idea of using model checking algorithms as a basis for SMT solving.

\bibliographystyle{splncs03}
\bibliography{qzy}

\end{document}